\documentstyle[prb,preprint,aps,psfig]{revtex}    % preprint form 
\begin{document} 
\draft 
\newcommand{\no}{\nonumber} 
\newcommand{\beq}{\begin{equation}} 
\newcommand{\eeq}{\end{equation}} 
\newcommand{\beqa}{\begin{eqnarray}} 
\newcommand{\eeqa}{\end{eqnarray}}
 
\title{Transport in normal-superconductor-normal structures \\ 
with local conservation of current}

\author{J. S\'anchez-Ca\~nizares and F. Sols} 
\bigskip 
\address{ 
Departamento de F\'{\i}sica Te\'orica de la Materia Condensada, C-V, 
and\\ 
Instituto Universitario de Ciencia de Materiales ``Nicol\'as 
Cabrera''\\ 
Universidad Aut\'onoma de Madrid, E-28049 Madrid, Spain}

\maketitle 
\begin{abstract} 

We study the transport properties of a NSN structure with an
insulating barrier at each NS interface. Coherent quasiparticle
scattering is assumed and self-consistency is implemented exactly to
guarantee local charge conservation. The presence of a finite
condensate flow has a greater influence on the transport properties
than either the gap depression near the interfaces or the coherent nature of
scattering. We find that a nonzero phase gradient causes a shift
towards lower voltages of the first peak in the differential
conductance and a global enhancement of current. At low currents, we
obtain gap profiles near the interfaces that are consistent with the
criteria for boundary conditions employed in macroscopic
descriptions. The existence of coherent multiple scattering gives
rise to a rich structure of resonances that is smoothed out for long
superconductors.

\end{abstract} 

\vspace{.5cm} 
\narrowtext 
 
\section{Introduction}

Motivated by the recent development of mesoscopic superconductivity,
\cite{hekk95}  the study of transport in normal-superconductor
structures has been the object of  renewed attention. Microscopic
studies are conventionally performed within the  framework of the  
Bogoliubov - de Gennes (BdG) equations: \cite{dege66}
\beqa 
\left[ \begin{array}{cc} 
H_0 & \Delta \\ 
\Delta^* & -H_0^* \end{array} \right] 
\left[ \begin{array}{c} u_{n} \\ 
v_{n} \end{array} \right] 
= \varepsilon_{n} 
\left[ \begin{array}{c} u_{n} \\ 
v_{n} \end{array} \right], 
\eeqa 
where $[u_n,v_n]$ and $\varepsilon_n$ are the wave function  and
energy of quasiparticle $n$. $H_0$ is the one-electron Hamiltonian
referred to the  Fermi energy and $\Delta$ is the pair potential. The
BdG equations are based on a mean  field Hamiltonian which does not
commute with the particle number  operator.\cite{dege66} It has been
noted \cite{furu91,bagw94,sols94,ferr90} that current  conservation
is only guaranteed if the BdG equations are solved within a
self-consistent scheme,  i.e., one in which  the gap function
$\Delta({\bf r})$  is required to satisfy the condition 
\beq 
\Delta = g \sum_{n} u_{n} v_{n}^* 
(1 - 2 f_{n}), 
\eeq 
$g$ being the electron-phonon coupling constant and $f_n$ the
occupation probability. 

The requirement of current  conservation can be satisfied only if the
condensate carries  a finite amount of current, \cite{sanc95,sanc97}
and for this a nonzero phase gradient  is needed:  $\nabla \varphi
\neq 0$, with $\Delta  \equiv |\Delta| e^{i\varphi}$.  At distances
from the scattering region much bigger than the coherence length,
the superfluid velocity $v_s \equiv \hbar \nabla \varphi/2m$ and the
gap amplitude acquire uniform values. Based on this result, a model
was developed in Ref.  \cite{sanc97} to compute the current-voltage 
characteristics for several normal-superconductor structures with the
simplifying assumptions of asymptotic  self-consistency (current is
globally conserved) and incoherent multiple scattering (if  more than
one interface is present). A rich transport behavior was unveiled of
which the  most salient features are a shift towards lower voltages
of the first peak in the  differential conductance and an enhancement
of current at not very low voltages caused  by the greater
availability of current carrying scattering channels. Different
voltage  regimes appear in the presence of a finite condensate flow,
since this creates a distortion  in the quasiparticle dispersion
relation within the superconductor which makes the  various
quasiparticle channels open at different voltages. A regime was
found in which  quasiparticles enter and leave the superconductor
only through Andreev transmission  and in which current is thus
insensitive to the presence of impurities within the superconductor. 
\cite{sanc96} It was also shown that a state of anisotropic gapless
superconductivity can  develop for high enough voltages if the
temperature is low.\cite{sanc95,sanc97} Effects  due to a finite
superfluid velocity are important when the supercurrent is 
comparable to the bulk critical current. This may occur already at
voltages of order  $\Delta_0/e$ if the interfaces are transmissive
and the superconductor is at most as  wide as the semi-infinite
normal leads, i.e., if there is no geometrical current dilution in
the  superconductor.\cite{sanc97,ried96} The self-consistent gap near
a NS interface with zero current was  already studied by McMillan
\cite{mcmi68} and, more recently, by Bruder  \cite{brud90} in the
context of anisotropic superconductivity.

The purpose of this paper is to go beyond the approximations
introduced in the  model of Ref. \cite{sanc97} and study the
nonlinear transport of a NSN structure within  a picture in which
quasiparticles undergo multiple coherent scattering and current is 
locally conserved (self-consistency is implemented exactly). In
particular, we wish to  know which of the transport properties found
in Ref. \cite{sanc97} survive in a more accurate description. Like in
previous scattering studies of transport through NS 
structures,\cite{sanc97,blon82} we assume that an insulating barrier
may exist at the  interface contributing $H\delta(x)$ to the
one-electron potential $H_0$. The  dimensionless parameter $Z \equiv
m H/\hbar^2 k_F$ is a measure of the barrier scattering  strength,
since the one-electron reflection probability is $Z^2/(1+Z^2)$. We
are interested  in NSN structures which contain two insulating
barriers -one at each interface- of  strengths $Z_1$ and $Z_2$. By
performing a systematic study of the dependence of  transport
properties on the values of $Z_1$ and $Z_2$, as well as on the length
of the  superconductor, we extend the work of Martin and Lambert
\cite{mart95} where a  similar set of assumptions was employed. 

In an asymmetric NSN structure, the distribution of the potential
drop between the two  interfaces has to be determined
self-consistently, like in the asymptotic model of Ref. 
\cite{sanc97}. Here we determine in addition the exact profile of the
pair potential $\Delta(x)$. Consistently with previous
work,\cite{sanc97} we find that the position of the first  peak in
the differential conductance (FPDC) is shifted to lower voltages,
because of the existence of a nonzero phase gradient. This effect is
more important than the local depression of $\Delta(x)$ or the
coherent nature of quasiparticle scattering. Due to the coherent
multiple scattering, a structure of resonances develops which gives 
rise to an oscillatory behavior of $dI/dV$ as a function of voltage
and length (see Figs. 4  and 5). The results obtained from the model
of incoherent scattering and asymptotic  self-consistency  are
recovered qualitatively for structures where the superconductor is
much longer than  its coherence length. 

\section{The model} 

We wish to solve the BdG equations (1) self-consistently within the
context of a one-dimensional model. The resulting gap (2) will in 
general be space dependent. At a given energy, a scattering problem
has to be solved  which is defined by Eq. (1) with $H_0$ describing a
NSN structure with insulating  barriers at the interfaces. 
Quasiparticle scattering states are labeled by the incoming channel
$\alpha$. Thus,  index $\alpha$ indicates the lead (left or right)
and the type (electron or hole) of the  initial quasiparticle state.
We can exploit the fact that only wave vectors in a small interval
around $\pm k_F$ are  important, and {\it linearize} the one-electron
Hamiltonian around the two Fermi points. In momentum representation,
\beq
\frac{\hbar^2 k^2}{2m}\equiv \frac{\hbar^2}{2m}(\pm k_F+\kappa)^2
\simeq E_F \pm \frac{\hbar^2 k_F \kappa}{m}=-E_F
\pm \frac{\hbar^2 k_F}{m}k.
\eeq
Therefore, instead of $H_0=(\hbar^2 /2m)(-d^2/dx^2)
-E_F$, we can write 
\beq 
H_0\simeq \pm i \, \frac{\hbar^2 k_F}{m} \frac{d}{dx} - 2E_F, 
\eeq 
in the vicinity of $\mp k_F$. For the numerical resolution, the BdG
equations are discretized within the  superconductor S, so that only
a finite set of positions $x_i=iL/N$ (with $i=0,1,...,N$)  are
considered, with a uniform spacing $\delta=L/N$. Within the interval
$[x_i-\delta/2, x_i+\delta/2]$ the solution of Eq. (1) can be
written  as a sum of quasiparticle plane waves,  $\sum_{\beta}
t^i_{\beta\alpha}e^{ik^i_{\beta}x}[u^i_{\beta},v^i_{\beta}]$  which
are solutions for the  ``uniform'' gap $\Delta(x_i)$. This is a safe
procedure if $\delta$ is chosen much  smaller than the
superconducting coherence length $\xi_0=\hbar v_F/\pi \Delta_0$, 
which is the typical length scale for gap variations. At each point
and energy, there are four possible quasiparticle states  labeled by
$\beta$. The components at $i+1$ are obtained in terms of the
components at $i$ through a $4 \times 4 $ transfer matrix that is 
determined by wave function matching at $x=(x_i+x_{i+1})/2$. When
a linearization scheme is adopted, 
the $\pm k_F$  branches do not couple within
the superconductor (only Andreev reflection or normal transmission
can occur) and the transfer matrix at each point is
box-diagonalized into two $2 \times 2$ matrices.  Right at the
interfaces ($x=0$ and $x=L$)  normal reflection and thus branch
mixing occurs, and in those points the transfer matrix is  also
determined by standard matching techniques. The global transfer
matrix is obtained by compounding the individual transfer matrices.

Once the scattering quasiparticles have been calculated, the gap is
determined from Eq.  (2). Due to normal reflection at the interfaces,
a quasiparticle $n$ has in general  components from both $\pm k_F$
branches. This gives rise to irrelevant oscillations in  $\Delta(x)$
on the scale of the Fermi wavelength, which we smooth out  by
introducing {\it only intra-branch interference terms} in the gap
equation (2). Once self-consistency has been achieved, the resulting
description does conserve charge  and the electric current must be
uniform. 

Self-consistency is implemented as follows. The input parameter is
the difference in  chemical potentials between the quasiparticles
coming from the left and the right normal  leads. The desired output
is the resulting current. The pair potential profile $\Delta(x)$ and
the  chemical potential at the superconductor $\mu_S$ are treated on
the same footing, as  parameters that are progressively adjusted to
achieve convergence. In the first scattering  calculation, some guess
values for the $\Delta(x)$ profile and $\mu_S$ are introduced. From 
the global transfer matrix we obtain the quasiparticle scattering
states, and from them  the current in the normal leads. For the
following iteration, the gap at each point is obtained from Eq.  (2)
and the superconductor chemical potential is adjusted to make the
currents in both leads equal. Self-consistency is achieved  when
$\Delta(x)$ and $\mu_S$ converge and the current in the  two leads is
the same within tolerable errors.

The main simplifications that we have introduced are: (i) Use of a
one-dimensional model, which we  expect to yield qualitatively
correct physics in structures with planar interfaces where  mode
mixing is not important.\cite{ried96}(ii) The voltage drop takes
place entirely at the interfaces  and the chemical potential within
the superconductor is uniform.  (iii) There is no proximity effect
($g(x)=0$ in  the normal metal). A similar set of approximations 
were employed in Ref., \cite{mart95} with the difference that no
linearization procedure was implemented there. Relaxation of (ii) and
(iii) may yield interesting physics, but here we want to focus on 
the effect of a nonuniform, complex $\Delta(x)$.

\section{Results} 

There are two main ways in which local current conservation can
affect the transport  properties of a NSN structure. One is the
induction of a nonzero phase gradient, which  has already been
mentioned. The second effect has to do with the local depression of
the  gap near the interface, which is a direct consequence of
self-consistency that already occurs at $v_s=0$. From the work  of
Refs. \cite{sanc95,sanc97,mart95} we have some understanding of the
effect of flow  and now we wish the explore the effect of gap
depression at the interface. \cite{vans87} In Fig. 1 we  plot for
comparison the Andreev reflection probability as a function of the
quasiparticle  energy for gap profiles with a step-like structure
(solid lines) and with a local, smooth  depression shown in the inset
(dotted lines). The profiles have been  obtained by truncating (and
matching with a uniform gap) the self-consistent gap of a NSN
structure with zero current.  Thus, there is no flow effect in any of
the two cases. A similar study has been performed by van Son et al.
\cite{vans87} introducing phenomenological gap profiles with a
proximity effect.  The influence on scattering of the local gap
reduction is small in the cases of low and high $Z$, but it is 
appreciable for intermediate $Z$. At energies $E \alt \Delta_0$,
Andreev reflection  (AR) is enhanced for $Z=0.5$ because it takes
place in three distinct (albeit coherent)  steps: the electron is
first normally transmitted, then it is Andreev reflected as it finds
a higher  gap, and finally the emerging hole is normally transmitted.
AR at the smooth gap increase  takes place with probability very
close to unity. On the other hand, the two normal  transmission (NT)
processes are more probable than direct AR (since this requires the 
simultaneous transmission of two particles), which is the only
possibility when  $E < \Delta(x=0)$. Thus, by being decomposed into
three processes of higher  probability, AR takes place more easily.
The effect of gap depression is very small at  $Z=0.1$ because direct
AR already has a high probability. For high $Z$, the gap  depression
is small and thus it cannot have important consequences.

In Figs. 2 and 3, we plot the gap amplitude $|\Delta|$ as a function
of position and voltage for  two asymmetric NSN structures:
$(Z_1,Z_2)=(0.1,0.5)$ (Fig. 2) and $(2,0.5)$ (Fig.  3).  Here, $Z_1$
and $Z_2$ are the barrier parameters at $x=0$ and $x=L$, 
respectively. In both cases, $L=8\xi_0$. At low voltages, $|\Delta|$
is strengthened if the  barrier is thick, while it is more strongly
depressed near a transmissive interface. This is  in qualitative
agreement with the Ginzburg-Landau result for the boundary conditions
describing the  interface between a superconductor and an insulator
or a normal metal: \cite{dege66} At a transmissive NS interface, the
order parameter must be zero, while at the interface  with an
insulator (here represented by a high $Z$ interface), it is the
derivative of the  order parameter what must vanish. We see again
\cite{zapa96} that the physics  obtained from a self-consistent
description at zero temperature is qualitatively similar to  that
which is derived from the Ginzburg-Landau approximation near the
critical  temperature. As the voltage increases, it drops mostly at
the  more reflecting interface, where a greater gap reduction takes place
because of the stronger presence of quasiparticles. In fact, Fig. 2
shows the rather counterintuitive feature that $|\Delta|$ can even
become  smaller at the more reflecting interface, in marked contrast
with the zero voltage behavior.

Upper Fig. 4 shows the differential conductance $dI/dV$ as a function
of voltage for  several structures as obtained from different types
of calculations. Thick and solid lines  have been obtained with exact
and asymptotic (see Ref. \cite{sanc97}) self-consistency, 
respectively. The other three lines have been obtained from  uniform
phase calculations, in the spirit of Ref. \cite{blon82}: The dotted
(dashed) line corresponds to incoherent  (coherent) scattering, while
the long-dashed line results from a coherent scattering calculation 
performed with the self-consistent gap profile of the zero current
case:  $\Delta(x,V=0)$. In all cases, $Z_2=0.5$, $L=8\xi_0$, and
$T=2$ K have been taken,  with superconductor material parameters
corresponding to those of Pb ($T_c=7.2$  K). The lower part of Fig. 4
shows the relevant energies of the problem at the midpoint
$x=L/2=4\xi_0$ corresponding to  the exactly self-consistent
calculation: $|\Delta|$ (dotted), $\Delta_{\pm}= |\Delta| \pm  \hbar
q v_F$ (dashed), and the voltage difference at each interface
(solid).

The first peak in the differential conductance (FPDC) occurs when
$eV$ reaches  $\Delta_-$, and its position is very sensitive to the
existence of finite superconducting  flow. This is particularly clear
in the case of the symmetric ($Z_1=0.5$) NSN structure.  The lowering
of the FPDC is less marked in the $Z_1=0.1$ case because the
asymmetric  voltage drop tends to mask the effect. \cite{sanc97} 
Finally, for $Z_1=2$, the FPDC is barely  shifted, because of a
global reduction of current caused by poor transmitivitty. As
compared with calculations based on asymptotic self-consistency,
implementation of local self-consistency and coherent quasiparticle
scattering displaces the FPDC position only slightly, and tends to
increase the current.

The existence of coherent multiple reflection at the interfaces gives
rise to a  structure of resonances that is clearly appreciated in the
plot of $dI/dV$ vs. voltage. The  oscillatory behavior of the
differential conductance predicted by coherent scattering 
calculations contrast with the less structured curves obtained from
the assumption of  incoherent scattering. These resonances are very
efficient at carrying current because  they are made of quasi-bound
states in which the quasiparticles are mostly Andreev  reflected at
the interfaces. Thus, for instance, a right-moving electron is
Andreev  reflected as a left-moving hole, which in turn is reflected
as a right-moving electron,  and repetition of the process results in
a strong net electron current to the right. These  oscillations are
very similar in nature to the Tomasch oscillations occurring in
tunnel junctions, \cite{toma65} and also occur in other transport
contexts involving coherent Andreev reflection. \cite{alls94}

The period of the oscillations in $dI/dV$ is strongly sensitive to
the  total superconductor length $L$. Actually, for a superconductor
with constant $\Delta=\Delta_0$, one would expect to have resonances
at energies
\beq
E_n=\Delta_0\left[1+\frac{(n-1/2)^2\pi^4\xi_0^2}{L^2}\right]^{1/2},
\eeq
which is obtained from requiring $k_+-k_-=(2n-1)\pi/L$ ($k_{\pm}$ are
the two possible solutions to $\varepsilon (k)=E$ in the $+k_F$
branch, $\varepsilon (k)$ being the quasiparticle dispersion
relation). In our case, Eq. (5) is not exactly fulfilled 
because of the effect of $v_s \neq 0$ on the pair potential.
Nevertheless, Fig. 5 shows that the separation between peaks 
decreases with increasing $L$, with a first peak that is not
very sensitive to the length  (see inset). The spacing between peaks
depends also on the values of $Z$ (not shown)  because of the effect
that these have on the effective gap amplitude.

\section{Conclusions}

We have performed a self-consistent scattering calculation of
nonlinear transport  through asymmetric NSN structures. By
introducing local current conservation and  coherent scattering of
quasiparticles, we have gone beyond the model employed in Refs. 
\cite{sanc95,sanc97}, where incoherent multiple scattering and global
current  conservation was assumed. Comparison of the results obtained
from the two different  models allows us to identify the most robust
physical features which are likely to  survive in realistic
scenarios. In both models we have obtained a lowering of the voltage 
threshold  for quasiparticle transmission (as signaled by the first
peak in the differential  conductance) and a global enhancement of
current caused by the increased availability  of charge-transmitting
scattering channels. Both effects are directly related to the
presence of a nonzero superfluid velocity in the condensate.

In general, the self-consistent gap does not have a uniform
amplitude.  Linearization of the BdG equations around the two Fermi
wave vectors automatically  washes out any spurious spatial
dependence on the scale of the Fermi wavelength, and  only the
relevant physics occuring at the scale of the coherence length is
preserved. The  gap amplitude $|\Delta(x)|$ tends to decrease with
increasing voltage and is  locally depressed in the vicinity of the
interface with the normal lead. In the presence of  a strongly
reflecting insulating barrier the depression is weak and $|\Delta|$ 
drops abruptly to zero at the interface (no proximity effect is
assumed). By contrast, if  the interface is transmissive, the gap
amplitude drops smoothly from its bulk value to a  very small value
right at the interface. This result is consistent with the criterion
for boundary conditions employed in macroscopic, Ginzburg-Landau
descriptions. \cite{dege66} However, we have found that  the
presence of a nonequilibrium population of quasiparticles may alter
this picture qualitatively.

Quasiparticle multiple scattering is responsible for the existence
of  resonances within the superconductor. These quasibound states are
very efficient current  carriers, since they consist mostly of
electrons that are Andreev reflected at the  interfaces, or
viceversa. Whenever the voltage reaches a threshold level that
permits  transmission through a new resonance, there is a quick rise
in the current, hence the  nonmonotonous dependence of the
differential conductance on the applied voltage (see  Figs. 4 and 5).
We have found that agreement with incoherent scattering calculations 
tends to improve as the superconductor length becomes large compared 
with the coherence length, because in that case  the structure of
resonances is smoothed out.

We end this article by pointing out some remaining challenges that
should be  addressed in future scattering studies of current
conserving transport. Self-consistency  should be extended to include
the Coulomb interaction term, since voltage variations  due to charge
pileup at the barriers may be comparable to the gap  if the interface
is appreciably reflecting. Most important, it  would be desirable to
understand how the transport properties discussed here and in 
previous references are affected by the presence of many modes which
can mix at the  interface and by collisions with impurities in the
superconductor. Finally, it will be of  interest to understand how
transport is modified by the existence of a proximity effect in  the
normal lead. 
	
\acknowledgments 
We wish to thank F. Beltram, C.J. Lambert, and A. Martin  for
valuable discussions.  This project  has been supported by
Direcci\'on General  de Investigaci\'on Cient\'{\i}fica  y T\'ecnica,
Project no. PB93-1248, and by the HCM and TMR  Programmes of the EU.
One of us (J.S.C.) acknowledges the support from Ministerio de 
Educaci\'on y Ciencia.\\

%Figures

%fig1
\begin{figure}[a]
\vspace{2cm}
\psfig{figure=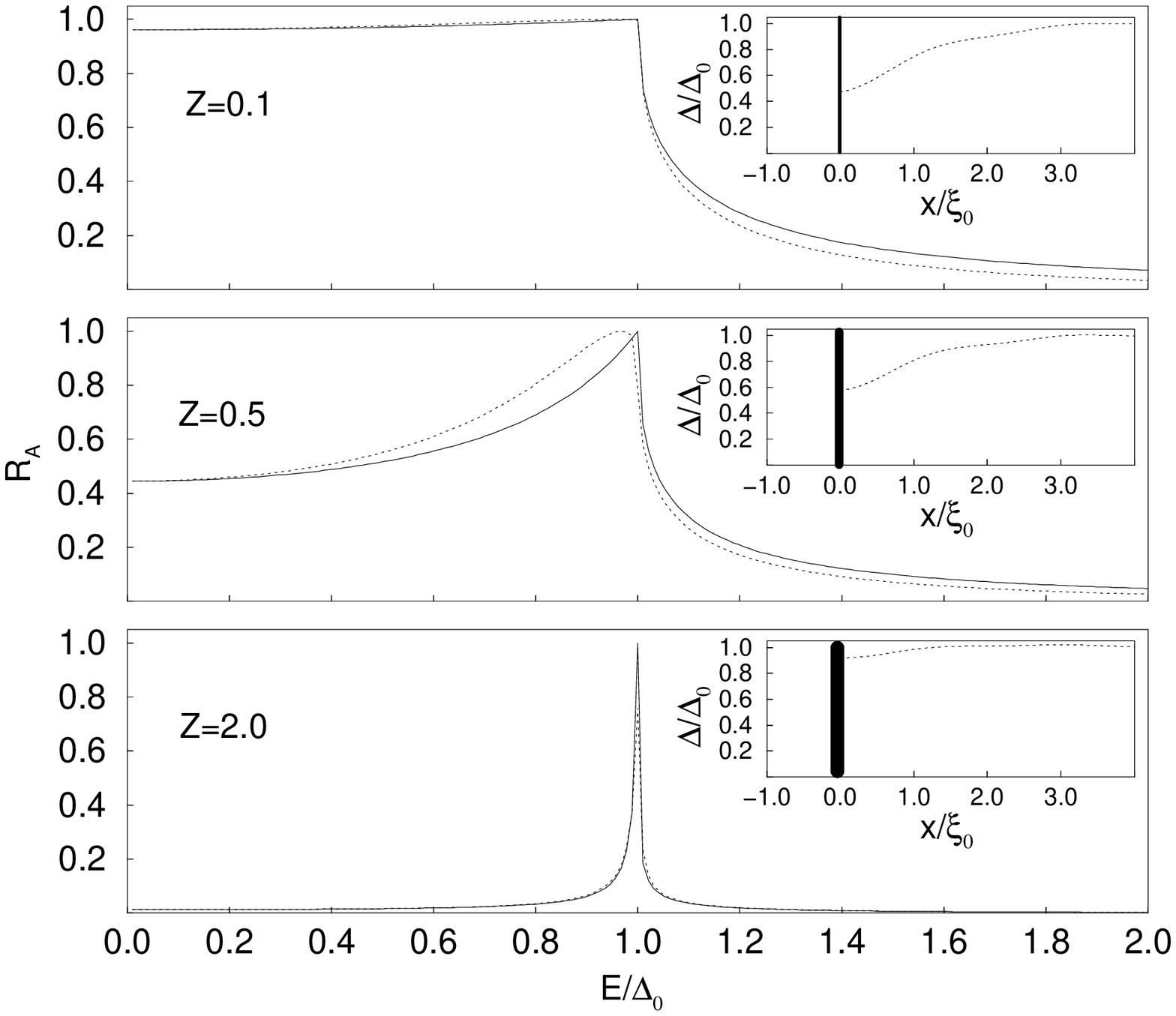,height=18cm,width=15cm}
\caption{ 
Andreev reflection at a NS interface with different insulating
barriers  for a stepwise (solid line) and self-consistent (dotted
line) pair potential. Insets show the self-consistent pair potentials
employed in the calculations.
} 
\end{figure}
 
%fig 2
\begin{figure}[b]
\hspace{-2.3cm}
\psfig{figure=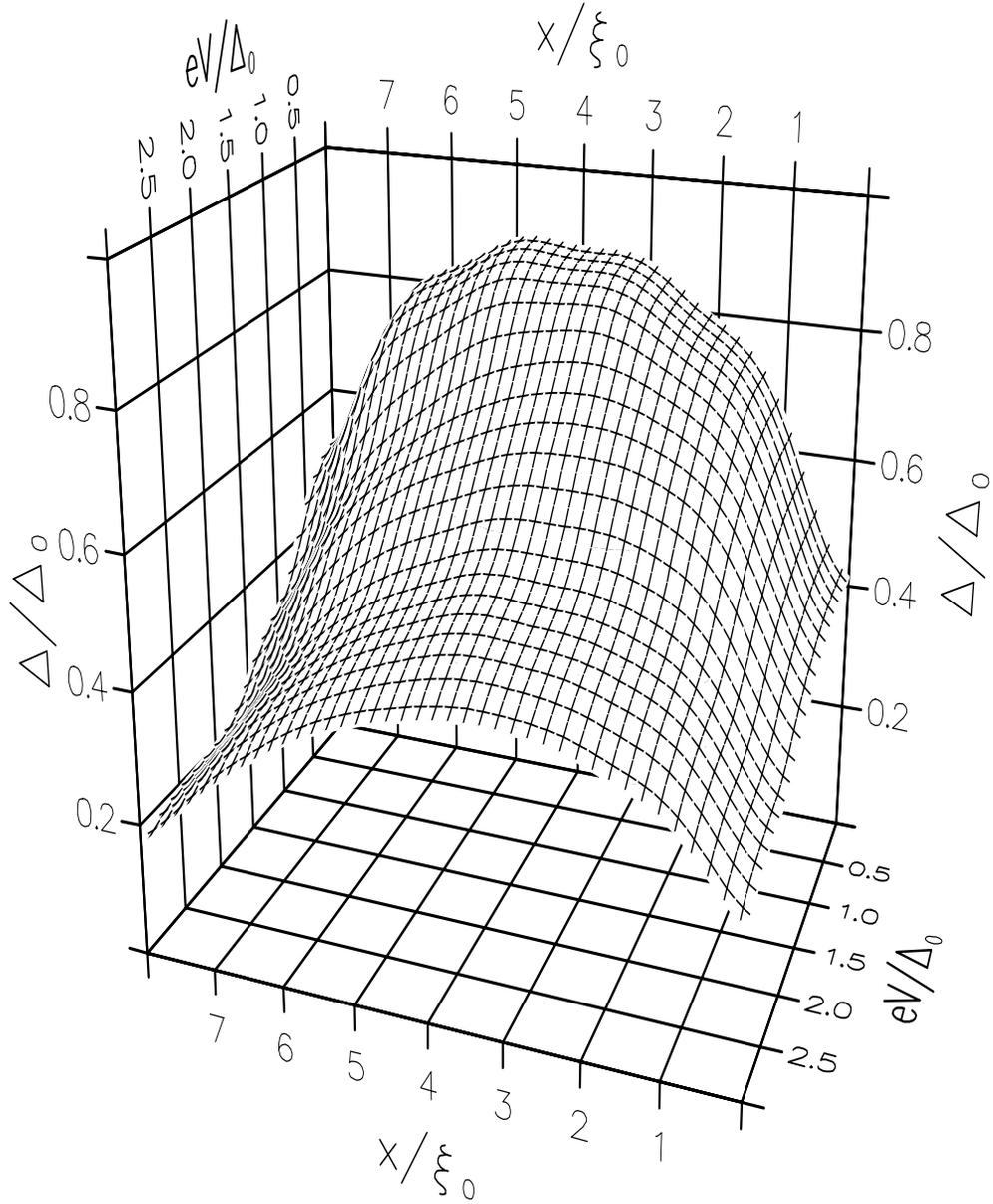,height=20cm,width=18cm,angle=90}
\caption{ 
Amplitude of the self-consistent pair potential for a NSN structure
with $Z_1=0.1$ (at $x=0$) and $Z_2=0.5$ (at $x=8\xi_0$) as a function
of position and applied voltage. A temperature of $T=2$ K (for
$T_c=7.2$ K) has been taken. 
} 
\end{figure} 
 
%fig 3
\begin{figure}[c]
\hspace{-2.5cm}
\psfig{figure=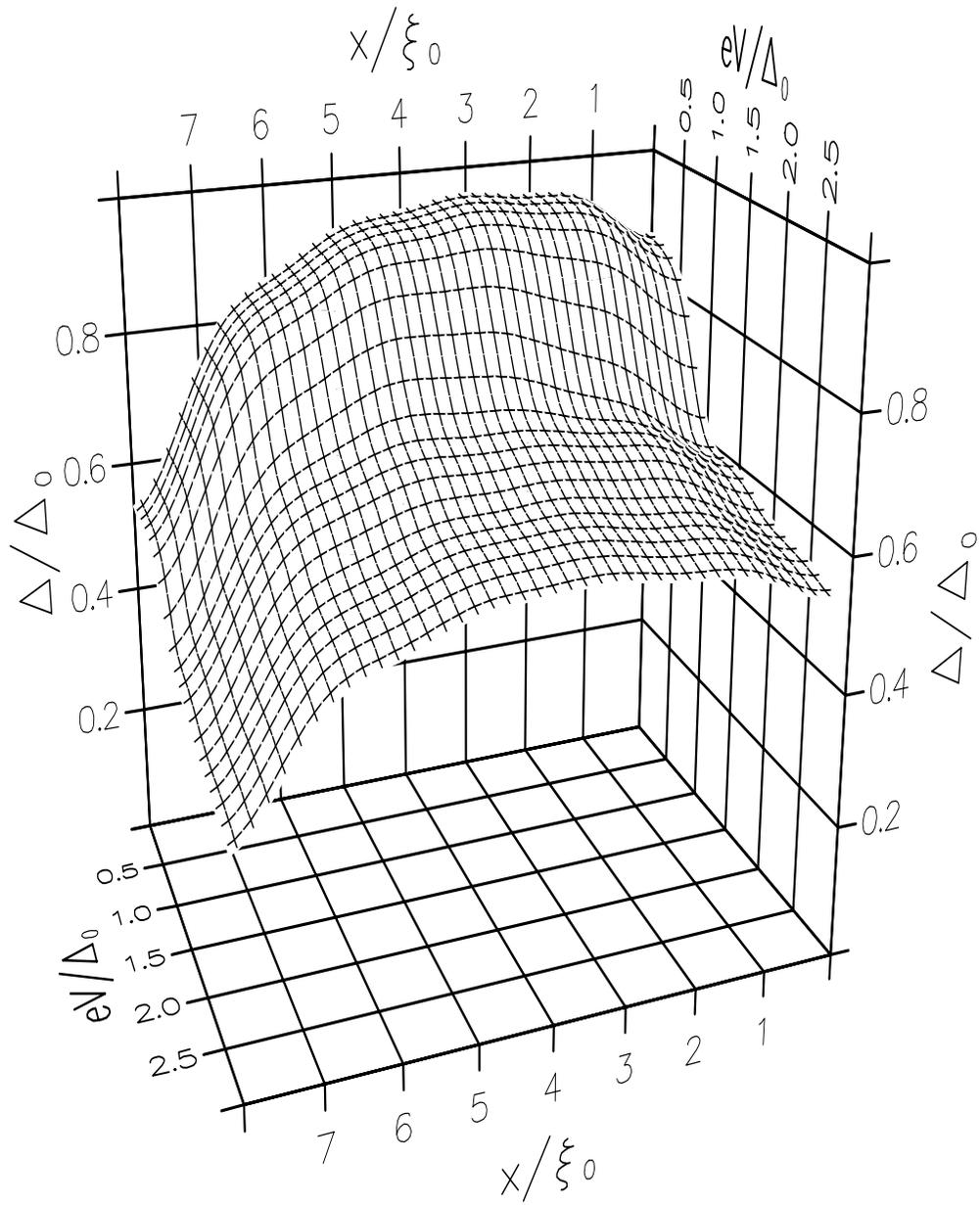,height=20cm,width=18cm,angle=90}
\caption{ 
Same as Fig. 2 for $Z_1=2$ and $Z_2=0.5$. 
} 
\end{figure} 
 
%fig 4
\begin{figure}[d]
\psfig{figure=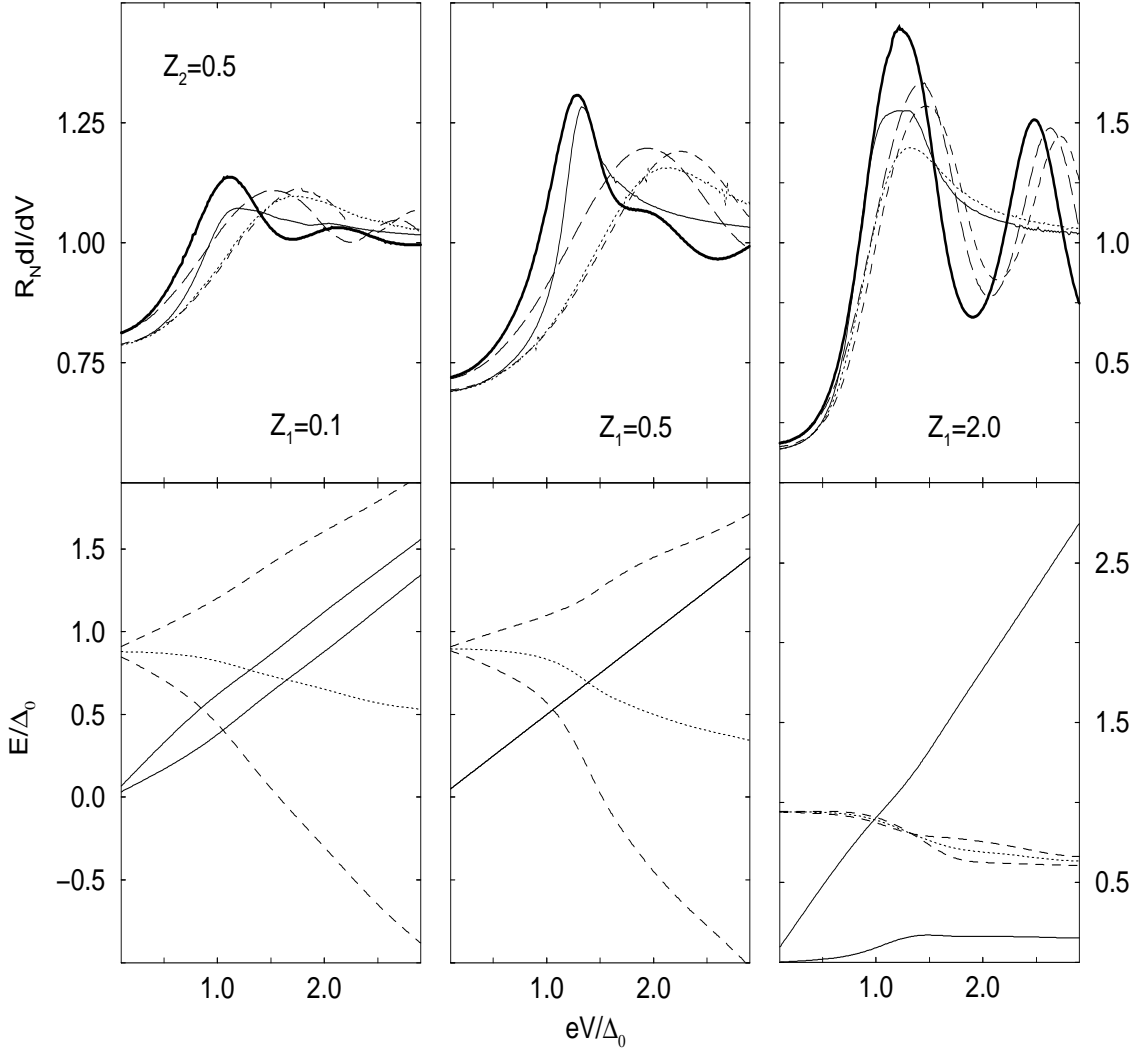,height=15cm,width=15cm,angle=270}
\caption{ 
NSN structure. $Z_2=0.5$ and data of Pb for S have been taken in all
cases. Temperature is $T=2$ K while $T_c=7.2$ K. The upper curves
show  the differential conductance calculated with exact (thick
solid line) and asymptotic  self-consistency (thin solid). The other
three lines result from calculations with a zero phase gradient:
Dotted (short-dashed) line corresponds to incoherent (coherent)
scattering with a stepwise pair potential, while the long-dashed line
corresponds to a coherent scattering calculation performed with a
zero-current gap. The lower curves show the relevant energies (in
units of $\Delta_0$) of the  problem in the exact calculation:
Voltage drops at each interface (solid),  magnitude of the order
parameter $|\Delta|$ (dotted),  and of the $\Delta_+$ and $\Delta_-$
thresholds (dashed) evaluated at $x=L/2=4\xi_0$.  
} 
\end{figure} 
 
%fig 5
\begin{figure}[e]
\psfig{figure=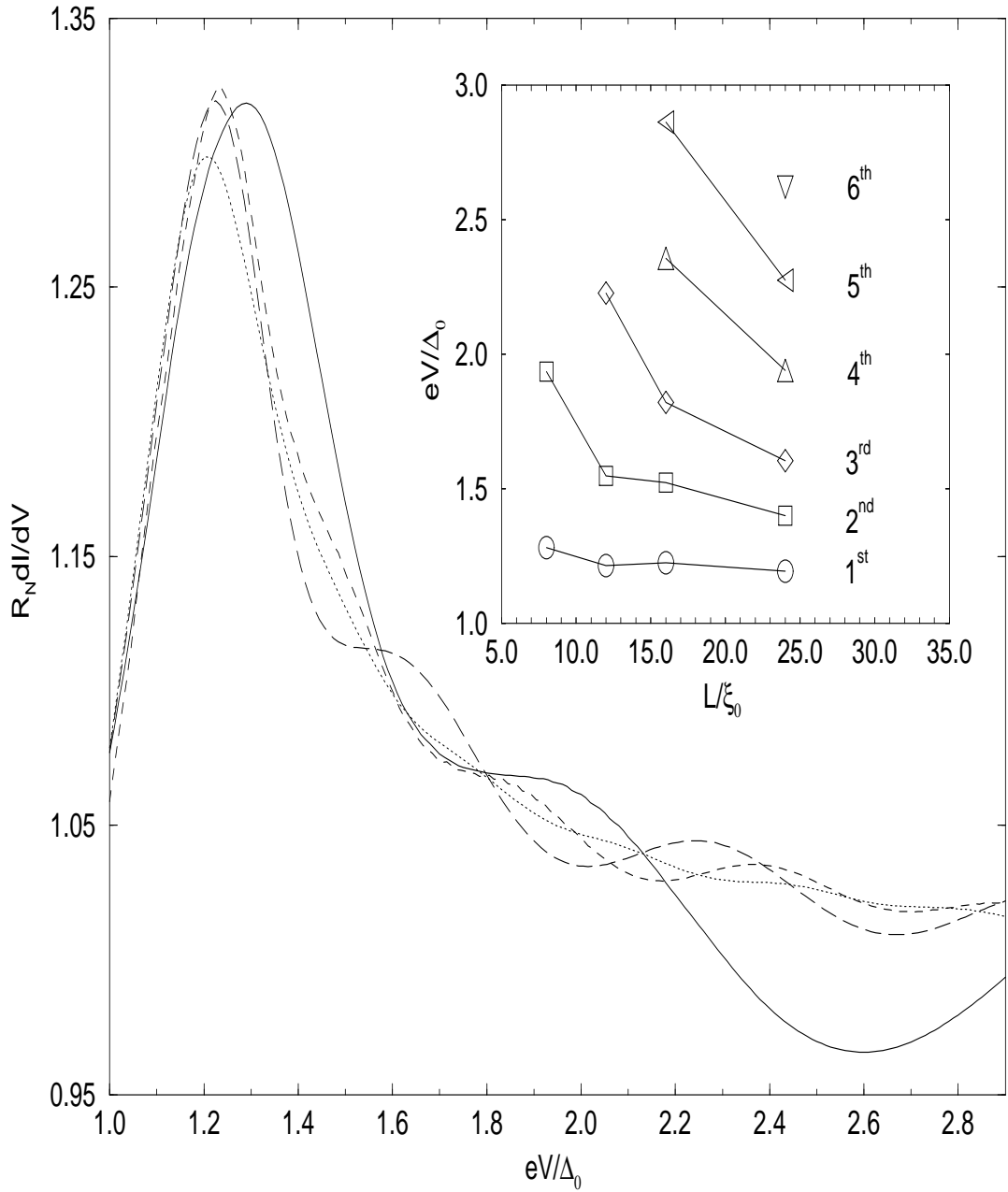,height=18cm,width=15cm,angle=270}
\caption{
{\it dI/dV} vs. {\it V} for the $Z_1=0.5$ case of Fig. 4 for
different superconductor lengths: $L=8\xi_0$ (solid), $L=12\xi_0$
(long-dashed), $L=16\xi_0$ (short-dashed), and  $L=24\xi_0$ (dotted).
Inset: Positions of the various resonance peaks as a function of the
superconductor length. 
} 
\end{figure}
         
\end{document}